\documentclass[12pt,a4paper]{article}
\usepackage{amsmath}
\usepackage{amssymb}
\usepackage{enumerate}
\usepackage{amsfonts}
\usepackage[latin1]{inputenc}
\begin{document}

%
\newcounter{saveeqn}
\newcommand{\alpheqn}{\setcounter{saveeqn}{\value{equation}}%
\setcounter{equation}{0}%
\addtocounter{saveeqn}{1}
\renewcommand{\theequation}{\mbox{\arabic{saveeqn}\alph{equation}}}}
\newcommand{\reseteqn}{\setcounter{equation}{\value{saveeqn}}%
\renewcommand{\theequation}{\arabic{equation}}}
\begin{center}

{\bf ``Vector Schwinger Model with a Photon Mass Term on the Light-Front'' \footnote{``Invited Contributed Talk'' delivered at the International Conference on ``Light Cone 2008: Relativistic  Nuclear and Particle Physics (LC2008)'', Mulhouse, France, July 07-11, 2008.}}
\\[35mm]
Usha Kulshreshtha\\
\vskip1cm
Department of Physics, Kirori Mal College\\
 University of Delhi, Delhi-110007, India \\
 Email: ushakulsh@gmail.com\\

\end{center}

\vspace{4cm}

\begin{abstract}
Vector Schwinger model with a mass term for the photon, describing 2D electrodynamics with massless fermions, studied by us recently, represents a new class of models. This theory becomes gauge-invariant when studied on the light-front. This is in contrast to the instant-form theory which is gauge-noninvariant. We quantize this theory on the light-front.
  
\end{abstract}

\newpage

The vector Schwinger model (VSM) describing two-dimensional electrodynamics with massless fermions\cite{1,2,3}, where the left-handed and right-handed fermions are coupled to the electromagnetic field with equal couplings, is of wide interest\cite{1,2,3} and the solutions of the theory have been obtained by several authors in various contexts. This model is characterized for its exact solvability, a property which is ensured by a remarkable feature of one-dimensional fermion systems, namely that they could be described in terms of canonical one-dimensional boson fields. This fermion-boson equivalence has led to the discovery of many interesting features of the two-dimensional field theories\cite{1,2,3}. Recently, we have studied this model with a mass term for the $ U(1)$ gauge field and studied its Hamiltonian and path integral formulations\cite{3,4} and the operator solutions in the usual instant-form of dynamics on the hyperplanes: $ x^{0} = t = $ constant. This modified model is seen to be gauge-noninvariant (GNI)in the IF of dynamics. It is important to emphasise here that although this model, has been studied in the literature rather widely but only without a photon mass term (which was a consequence of demanding the regularization  for the VSM to be gauge-invariant(GI)\cite{1}.

This theory represents a new class of models in the two-dimensional quantum electrodynamics with massless fermions but with a photon mass term. However, as expected this theory is seen to be a gauge-noninvariant (GNI) theory owing to the presence of a mass term for the vector gauge field and a corresponding presence of a set of second-class constraints in the theory. When this theory is studied on the light-front using the front-form (FF) of dynamics\cite{5}, as is done in the present work, the theory is seen to be a gauge-invariant (GI) theory possessing a set of first-class constraints. In the present work we quantize this theory on the light-front (i.e., on the hyperplanes defined by the equal light-cone time $\tau = x^{+} = (x^0 + x^1)/\sqrt{2} = $ constant \cite{5}. After a very brief recapitulation of the Schwinger model, we define the modified theory which involves a mass term for the $U(1)$ gauge field $ A^\mu (x) $\cite{3}.
 
We now start with the generalized Schwinger model (GSM) which  describes the quantum electrodynamics in one-space one-time dimension with massless fermions and contains both: the VSM as well as the CSM,  defined by the action\cite{1}:
\alpheqn
\begin{eqnarray}
{S}_{1} &=& \int_{ }^{} {\cal L}_{1} (\psi , \bar\psi , A^{\mu}) d^{2}x \\
\label{1a}
{\cal L}_{1}&=& \biggl[ i \bar{\psi} \gamma^\mu \partial_\mu \psi + \frac{1}{2} e_{R} \bar{\psi} \gamma^\mu ( 1 + \gamma^{5} ) \psi A_\mu  + \frac{1}{2} e_{L} \bar{\psi} \gamma^\mu ( 1 - \gamma^{5} ) \psi A_\mu  - \frac{1}{4}F_{\mu\nu} F^{\mu\nu} \biggr] \\ 
\label{1b}
g_{1} &=& \frac{1}{2} ( e_{L} - e_{R} ) \qquad; \qquad g_{2} = \frac{1}{2} ( e_{L} + e_{R} ) \quad;\quad \gamma^{\mu}\gamma^{5} = - \epsilon^{\mu\nu}\gamma_{\nu} \\
\label{1c}
\gamma^{5} &=& \gamma^{0} \gamma^{1} =  \left( \begin{array}{ll} +1 & ~~0 \\ ~~0 &-1 \end{array} \right)\quad ; \quad
\epsilon^{\mu\nu} =\epsilon^{-\nu\mu} =  \left( \begin{array}{ll} ~~0 & +1 \\ -1 & ~~0 \end{array} \right) \\ 
\label{1d}
g^{\mu\nu}& := &  g_{\mu\nu} = diag(+1, -1)  \quad;\quad 
F^{\mu\nu} = ( \partial^\mu A^\nu  -  \partial^\nu A^\mu)\quad ;\quad \mu,\nu = 0,1
\label{1e}
\end{eqnarray}
\reseteqn
which is equivalent to its bosonized form: 
\alpheqn
\begin{eqnarray}
{S}_{2} &=& \int_{ }^{} {\cal L}_{2} (\phi , A^{\mu}) d^{2}x   \\
{\cal L}_{2} &=& \biggl[ \frac{1}{2}\partial_\mu \phi \partial^\mu \phi
 + ( g_{1} g^{\mu\nu} - g_{2} \epsilon^{\mu\nu} ) \partial_\mu \phi  A_\nu - \frac{1}{4}F_{\mu\nu} F^{\mu\nu} + \frac {M^2}{2\pi } A_{\mu} A^{\mu} \biggr] 
\label{2}
\end{eqnarray}
\reseteqn
Here, the mass term for $A_{\mu}$ arises from the regularization ambiguities associated with the definition of the current. Here the case of chiral Schwinger model (CSM) is obtained from the GSM by setting $g_{1} = g_{2} = g $ ( i.e., $e_{R} = 0 $); and  $M^2 = a g^{2} $ , where $ a $ is the regularization parameter in the so-called standard regularization \cite{1}. However, the case of vector Schwinger model (VSM) is obtained by setting $ g_{1} = 0 , g_{2} = e $ ( i.e., $ e_{L} = e_{R} =  e  $) ; and  $ M = 0 $. Here,  $ e_{L} = e_{R} $ ( $ =  e  $) implies a vector -like theory. Also in the case of VSM, demanding the regularization to be gauge-invariant (GI) fixes $ a = 0 $ i.e., $ M = 0 $ where as in CSM, no choice for the value of  $ a $   can make the theory GI and therefore  $a $ is left as a free parameter\cite{1,2,3,4,5}. The VSM more commonly called as the Schwinger model is defined by the action\cite{1}:
\alpheqn
\begin{eqnarray}
{S}_{3} &=& \int_{ }^{} {\cal L}_{3} (\psi , \bar\psi , A^{\mu}) d^{2}x   \\
{\cal L}_{3} &=& \biggl[ \bar\psi \gamma^\mu (i\partial_\mu + e A_\mu ) \psi - \frac{1}{4}F_{\mu\nu} F^{\mu\nu} \biggr]
\label{3}
\end{eqnarray}
\reseteqn
which is equivalent to its bosonized form: 
\alpheqn
\begin{eqnarray}
{S}_{4} &=& \int_{ }^{} {\cal L}_{4} (\phi , A^{\mu}) d^{2}x  \\ 
{\cal L}_{4} &=& \biggl[ \frac{1}{2}\partial_\mu \phi \partial^\mu \phi
 - e \epsilon^{\mu\nu} \partial_\mu \phi  A_\nu - \frac{1}{4}F_{\mu\nu} F^{\mu\nu} \biggr] 
\label{4}
\end{eqnarray}
\reseteqn
where $e$  is the coupling constant that couples the massless fermion(or equivalently the boson) with the $ U(1)$ gauge field $ A^{\mu}  $. This theory is a well known gauge-invariant theory, possessing a set of two first-class constraints \cite{4}.  We now modify the above theory by including a mass term for the $U(1)$ gauge-field $A^\mu $, into the above Lagrangian density, defined by: $ [ {\cal L}_{m} = (a/2) e^2 A_\mu A^\mu ] $ , where $a$ is the standard regularization parameter\cite{1,2}.  The modified resulting theory then describes the Schwinger model with a photon mass term defined by the action:
\alpheqn
\begin{eqnarray}
S &=& \int_{ }^{} {\cal L }(\phi , A^{\mu}) d^{2}x \\
\label{5a}
{\cal L }&=& [ {\cal L}_{4} + {\cal L}_{m} ] \nonumber \\  
& = &   \biggl[ \frac{1}{2}\partial_\mu \phi \partial^\mu \phi - e \epsilon^{\mu\nu} \partial_\mu \phi  A_\nu - \frac{1}{4}F_{\mu\nu} F^{\mu\nu} 
 + \frac{1}{2} a e^2 A_\mu A^\mu  \biggr] 
\label{5b}
\end{eqnarray}
\reseteqn
The first term in the above action represents a massless boson, which is equivalent to a massless fermion in two-dimensions, the second term represents the vector coupling of this fermion to the electromagnetic field $A^{\mu} $, the third term is the kinetic energy term of the electromagnetic field and the fourth term is the mass term for this electromagnetic field. The theory defined by (5) when considered in the IF of dynamics is seen to possess one primary and one secondary Gauss-law constraint:
\begin{eqnarray}
\chi_{1} = \Pi^{0} \approx 0 \quad; \quad
\chi_{2} = [  \partial_{1} E + e \partial_{1} \phi + a e^2 A_{0} ] \approx 0
\label{6}
\end{eqnarray}
where  $\pi $ , $\Pi^{0}$ and $E $(= $\Pi^{1}$), are the canonical momenta, conjugate respectively to  $\phi$ ,  $A_{0}$ and $A_{1}$. The matrix of the Poisson brackets among the constraints $\chi_{ 1}$ and $\chi_{ 2}$ is seen to be nonsingular implying that the set of constraints $\chi_{ 1}$ and  $\chi_{ 2}$ is second-class and that the theory under consideration is gauge-noninvariant (GNI). This  theory has been quantized in Ref.\cite {3}. We now consider this theory defined by (5) on the light-front i.e.,  on the hyperplanes defined by the equal light-cone time $\tau = x^{+} = (x^0 + x^1)/\sqrt{2} = constant $. In the LF frame the canonical momenta $\pi $ , ${\Pi}^{+}$ and $ {\Pi}^{-} $, conjugate respectively to  $\phi$ ,  $A_{+}$ and $A_{-}$ obtained from the above action in the light-cone (LC) coordinates are:
\begin{equation}
\pi = (\partial_{-} \phi  +  e A^{+}) \quad;\quad
\Pi^{+} = 0 \quad;\quad
\Pi^{-} =  (\partial_{+} A^{+} - \partial_{-} A^{+} ) 
\label{7}
\end{equation}
The above equations imply that the theory possesses two primary constraints
\begin{equation}
\psi_{1} = \Pi^{+} \approx 0 \quad; \quad  \psi_{2} = ( \pi -   \partial_{-} \phi - e  A^{+} ) \approx 0 
\label{8}
\end{equation}
After including the primary constraint  $\chi_{1}$ in the canonical Hamiltonian density ${\cal H}_c$ with the help of Lagrange multiplier fields $ u(x,t)$ and $ v(x,t) $which are treated as dynamical, the total Hamiltonian density ${\cal H}_{T}$ could be written as:
\begin{equation}
{\cal H}_T = \biggl[ \frac{1}{2} {({\Pi}^{-})}^{2} + {\Pi}^{-} \partial_{-} A^{-}  + e A^{-}{\partial}_{-}{\phi}  - a {e}^{2} A^{+} A^{-} + u \Pi^{+} + v ( \pi -   \partial_{-} \phi - e  A^{+} )    \biggr]
\label{9}
\end{equation}
The Hamilton's equations of motion of the theory that preserve the constraints of the theory in the course of time could be obtained from the total Hamiltonian:$H_{T}= \int_{ }^{} {\cal H}_{T} dx $ and are omitted here for the sake of bravity. The theory is also seen to possess two secondary Gauss-law constraints: 
\begin{equation}
\psi_{3} = ({\partial}_{-}{\Pi}^{-} - e \partial_{-}\phi + a e^2 A^{+} ) \approx 0      \quad ; \quad  \psi_{4} =  a e^{2} {\Pi}^{-}
\label{10}
\end{equation}
The matrix of the Poisson brackets among the constraints $\psi_{ i}$  is seen to be singular implying that the set of constraints $\psi_{i} $is first-class and that the theory is GI. The theory is indeed seen to be invariant under the local vector gauge transformations:
\alpheqn
\begin{eqnarray}
\delta \phi &=&  - e \beta \quad;\quad \delta A^{+} = \partial_{-}\beta \quad;\quad   \delta A^{-}= \partial_{+}\beta  \quad;\quad 
 \delta u = \partial_{+}\partial_{+} \beta  \\
 \delta v &=& - e \partial_{+} \beta \quad;\quad \delta \pi = \delta \Pi^{+}  = \delta \Pi^{-} = \delta \Pi_{u}  = \delta \Pi_{v} =  0 
\label{11}
\end{eqnarray}
\reseteqn
where $\beta \equiv \beta $ ( $ \tau , x^{-} $ ) is an arbitrary function of its arguments. The vector gauge current of the theory  $J^\mu \equiv (J^{+} , J^{-}) $  is:
\alpheqn
\begin{eqnarray}
J^{+} &=& \int j^{+} dx  = \int  dx  \biggl[ - e \beta \partial_{-}\phi -  e^2 \beta A^{+} + \partial_{-}\beta (\partial_{+} A^{+} - \partial_{-} A^{-}) \biggr] \\
J^{-} &=& \int j^{-} dx   = \int  dx  \bigg[ - e \beta \partial_{+}\phi +  e^2 \beta A^{-} - \partial_{+}\beta (\partial_{+} A^{+} - \partial_{-} A^{-}) \biggr]
\label{12}
\end{eqnarray}
\reseteqn
The divergence of the vector gauge current density of the theory vanishes (giving $ \partial_{\mu} j^{\mu} $ = $  0  $), implying that the theory possesses at the classical level, a local vector-gauge symmetry. Now the theory could be quantized using the standard Hamiltonian and path integral quantization procedures e.g., under the gauge-fixing conditions or the gauge constraints:  
\begin{eqnarray}
\eta_{1} =  A^{+} \approx 0 \quad;\quad \eta_{2} =  A^{-} \approx 0
\label{13}
\end{eqnarray}
This LF theory could now be easily quantized using the standard Dirac quantization procedure. The results are omitted here for the sake of brevity. While considering the path integral formulation, the transition to quantum theory is made again by writing the vacuum to vacuum transition amplitude for the theory,  called the generating functional  $ Z[J_k] $ of the theory in the presence of the external sources: $J_k$ as follows\cite{4}:
\begin{eqnarray}
Z[J_{k}] &=& \int [d\mu] \exp \biggl [i \int d^{2} x \biggl [ J_{k} \Phi^{k} + \pi \partial_{+}\phi +  \Pi^{+}\partial _{+}A^{-} + \Pi^{-}\partial_{+}A ^{+}   \nonumber  \\
&& \qquad \qquad \qquad \qquad \qquad \qquad \quad \quad + \Pi_{u}\partial_{+}u  + \Pi_{v}\partial_{+}v  - {\cal H}_{T} \biggr ] \biggr ]  
\label{14}
\end{eqnarray}
where the phase space variables of the theory are: $\Phi^{k} \equiv (\phi, A^{-}, A^{+}, u, v)$ with the corresponding respective canonical conjugate momenta: $\Pi_{k} \equiv (\pi, \Pi^{+}, \Pi^{-}, \Pi_{u}, \Pi_{v}) $. The functional measure $[d\mu]$ of the generating functional $Z[ J_{k}]$ under the above LC gauges is obtained as\cite{4}: 
\begin{eqnarray}
[d\mu] &=&  [  [\partial_{-}\delta(x^{-} - y^{-})] [\delta(x^{-} - y^{-})] ] [d\phi] [dA^{+}] [dA^{-}] [du] [dv]         \nonumber \\
& & \qquad [d\pi] [d\Pi^{-}]  [d\Pi^{+}] [d\Pi_{u}] [d\Pi_{v}] \nonumber \\
& & \qquad  \delta [ \Pi^{+}  \approx 0]  \delta [ (\pi - \partial_{-} \phi - e A^{+}) \approx 0 ]  \nonumber \\
& & \qquad \delta [ (\partial _{-}\Pi^{-} - e \partial _{-} \phi + a e^2 A^{+}) \approx 0 ] 
\nonumber \\
& & \qquad \delta [ a e^2 \Pi^{-} \approx 0 ] [\delta [A^{+} \approx 0] \delta [ A^{-} \approx 0] 
\label{15}
\end{eqnarray}
This completes the Hamiltonian and path integral formulations of our LF theory.

\end{document}